\documentclass[twocolumn,showpacs,preprintnumbers,prb]{revtex4}
\usepackage{dcolumn}
\usepackage{amssymb}
\usepackage{bm}
\usepackage{graphicx}
\usepackage{float}
\usepackage{subfigure}

\newcommand{\ket}[1]{\left| #1 \right>} 
\newcommand{\bra}[1]{\left< #1 \right|} 

\begin{document}

\title{Entanglement spectrum and entangled modes of highly excited states in random XX spin chains}
\author{Mohammad Pouranvari and Kun Yang}
\affiliation{National High Magnetic Field Laboratory and Department of Physics,
Florida State University, Tallahassee, Florida 32306, USA}
\date{\today}

\begin{abstract}
We examine the real space renormalization group method of finding \textit{excited eigenstate} (RSRG-X) of the XX spin-1/2 chain, from
entanglement perspectives. Eigenmodes of entanglement Hamiltonian, especially the maximally entangled mode and corresponding entanglement energies are studied and
compared with predictions of RSRG-X. Our numerical results demonstrate the accuracy of the RSRG-X method in the strong disorder limit, and quantify its error when
applied to weak disorder regime.
Overall, our results validate the RSRG-X method qualitatively, but also show that its accuracy decreases with increasing temperature.
\end{abstract}

\maketitle

\section{Introduction}
Entanglement has emerged as an increasingly important way to characterize phases and phase transitions in quantum many-particle systems\cite{zeng}. Among different
measures of entanglement, von Neumann entanglement entropy (EE) remains the most popular one, and it is defined as follows. If a system is in a pure state
$\ket{\Psi}$, the density
matrix of the system is $\rho=\ket{\Psi}\bra{\Psi}$; EE is the von Neumann entropy of the reduced density matrix (RDM) of a chosen subsystem, which in turn is
obtained
by tracing over degrees of freedom outside of the subsystem. Since RDM is a positive definite operator, we can write the RDM of subsystem $A$, $\rho^A$ as  exponential
of a Hermitian operator,
$H^A$:
\begin{equation} \label{eq:rho}
\rho^{A}=\frac{1}{Z} e^{-H^{A}},
\end{equation}
where $Z=Tr (e^{-H^A})$, and we call $H^A$ entanglement Hamiltonian of subsystem $A$. Entanglement spectrum (ES), which is the spectrum of $H^{A}$, has also been found
to be highly valuable\cite{ref:lihaldane}.

In addition to the EE and ES, we recently studied eigenmodes of entanglement Hamiltonian.
For the special case of free fermion system (that we consider in this and earlier papers), the entanglement Hamiltonian $H^A$ is also a {\em free fermion}
Hamiltonian:\cite{rdm}
\begin{equation} \label{eq:HA}
H^{A}=\sum _{i,j=1}^{N_A} h_{ij}^A c_{i}^{\dagger} c_{j}.
\end{equation}
We find the single-particle eigenmode that contributes most to EE, the maximally
entangled mode (MEM) (see below for definition), contains rich physical information about the system \cite{ref:pouranvariyang1,ref:pouranvariyang2}. In Ref.
\onlinecite{ref:pouranvariyang1}, we studied the \textit{ground state} entanglement properties of random anti-ferromagnetic XX spin chain and compared it with the
predictions
of the real space renormalization group (RSRG) method \cite{ref:dsfisher}. At each step of RSRG method, two spins form a singlet and the ground state is approximated
by tensor product of singlets formed with all lengths. Expectations of RSRG about entanglement are that MEM be highly localized at the location of two spins that are
entangled and also
entanglement energies be either $0$ for such entangled spins that cross the boundary, or $\pm \infty$ otherwise. Our study showed that these predictions of RSRG on
ground state are indeed correct in the strong disorder regime.
Furthermore, in the weak disorder regime, where RSRG remains qualitatively correct at long length scales but quantitatively inaccurate at short distances, MEM gives
the profiles of the {\em effective} spins that form singlets.

Recently more attention has been focused on entanglement entropy of  \textit{highly excited states} (namely states with finite excitation energy {\em density}),
including free fermion systems \cite{ref:huangmoore, ref:Parameswaran, storms, laiyang}.
In the meantime, a new study extended the idea of RSRG to include calculation of excited eigenstate of random spin chains, which is named RSRG-X method
\cite{ref:rsrgx}. This
method relies on the fact that at each step of eliminating spins coupled by the strongest bond, one may choose to put them in the ground state or one of the excited
states of the bond. For example, in the case of XX spin chain, either singlet or each one of the triplet states can be chosen for two spins, based on probabilities
given by the temperature. Repeating this procedure generates a typical highly excited state for a given temperature. Justification of RSRG-X is subtler than RSRG, as
there exist distinctive excited states generated by RSRG-X whose energies are very close, which may be mixed by interactions that are left out in RSRG-X.

The main purpose of this paper is to critically examine the RSRG-X method using insights from entanglement, and establish its validity by comparing its predictions
about the entanglement properties of the system with exact numerical calculations. We use methods similar to Ref. \onlinecite{ref:pouranvariyang1}, with substantial
extensions needed to handle highly excited states. As detailed in the following sections, our results agree with the predictions of RSRG-X method qualitatively. We are
also able to quantify the error RSRG-X  makes, and show that generally speaking such error increases with increasing energy density/temperature, and decreasing
disorder strength.

This paper is organized as follows. In sec. \ref{sec:Modelandentangledmodes}, the model that we study and also terminology about EE that is used in this paper and
calculation methods are introduced. Sec. \ref{RSRG-X} is devoted to the explaining of the RSRG/RSRG-X method and applying it to our model. The main section is
\ref{results} which has two subsections, in one we study entangled modes and the other is about entanglement energy. Finally, a summary is given in the sec.
\ref{conclusion}.

\section{Model and Entangled Modes} {\label{sec:Modelandentangledmodes}}

The model we work with is a one-dimensional (1D) spin-1/2 XX model with $N$ sites and with random nearest neighbor couplings $J$. The Hamiltonian of the system is

\begin{equation} \label{eq:origham}
H=\sum_{n=1} ^{N-1} J_{n} (s_{n}^{x} s_{n+1}^{x}+s_{n}^{y}s_{n+1}^{y}) .
\end{equation}
We use open boundary condition. By mapping spin operators to fermion operators via Jordan-Wigner transformation:

\begin{equation}
c_{n}=e^{i \pi \sum_{j<n} s_{j}^{+}s_{j}^{-}} s_{n}^{-} ,
\end{equation}
where $c (c^{\dagger})$ is fermion annihilation (creation) operator, the fermionic representation of the Hamiltonian is:

\begin{equation} \label{eq:hamc}
H=\frac{1}{2} \sum_{n=1} ^{N-1} J_{n} (c_{n}^{\dagger} c_{n+1}+c_{n+1}^{\dagger}c_{n}).
\end{equation}
$J$'s in Eq. (\ref{eq:hamc}) will be generated based on random distribution functions to be specified later.
The number of fermions is related to magnetization as:
\begin{equation}\label{eq:nf}
N_{F}=\frac{1}{2} N + \sum_{n=1}^{N} s_{n} ^{z}.
\end{equation}

In this work we diagonalize Eq. (\ref{eq:hamc}) and generate its {\em exact} eigenstates, study their entanglement properties, and compare with predictions of RSRG-X.

To study bipartite EE, we divide the system into two subsystems $A$ and $B$, often (but not necessarily) with equal number of sites. Subsystem $A$ is from site $1$ to
$N_A$, and subsystem $B$ is from site $N_A+1$ to $N$. RDM of subsystem $A$ is written as Eq. (\ref{eq:rho}) which contains entanglement Hamiltonian written as Eq.
(\ref{eq:HA}). To determine eigenmodes and eigenvalues of $h^{A}$ we follow Ref. [\onlinecite{ref:correl}] by defining correlation function
\begin{equation}  \label{eq:cor}
C_{ij}^A=<c_{i}^{\dagger} c_{j}>,
\end{equation}
which is an $N_A \times N_A$ Hermitian matrix. Eigenmodes of the correlation function are the same as those of $h^A$, while eigenvalues of correlation function,
$n_k^A$, and eigenvalues of $h^A$, $\epsilon_k^A$, are related to each other:
\begin{equation} \label{eq:zeta}
n_k^A=\frac{1}{1+e^{\epsilon_k ^A}},
\end{equation}
where $\epsilon_k ^A$  referred to as entanglement energy, and $n_k ^A$ is the probability of finding a fermion in the corresponding mode. In terms of the $C^A$
eigenvalues, the entanglement entropy is:
\begin{equation}
\text{EE}=-\sum_{k=1} ^{N_F} [n_k^A \ln(n_k^A)+(1-n_k^A) \ln(1-n_k^A)].
\end{equation}

There is no contribution to the EE when $n_k^A=1$ ($n_k^A=0$) which corresponds to $\epsilon_k^A=-\infty$ ($\epsilon_k^A=+\infty$) and the most contributions comes
from $n_k^A$'s close to $1/2$ (i.e. $\epsilon_k^A$ be close to $0$).

RDM of subsystem $B$ can also be written as the exponential of a Hermitian operator similar to Eq. (\ref{eq:rho}), and also a correlation function for subsystem $B$
can be defined similar to Eq. (\ref{eq:cor}). Thus, entanglement Hamiltonian eigenmodes of both subsystem $A$ and $B$ can be obtained. We use the Klich
method\cite{ref:klich} to stick together eigenmodes of subsystem $A$ (which are from site $1$ to $N_A$) and eigenmodes of subsystem $B$ (which are from site $N_A+1$ to
$N$). Projecting  Klich eigenmodes to the subsystem $A$ ($B$) yields to the entanglement eigenmodes of the subsystem $A$ ($B$) weighted by the probability of finding a
fermion in those eigenmodes. For each Klich eigenmode, which is filled by one fermion, $n_k^A+n_k^B=1$ so $\epsilon_k^A=-\epsilon_k^B$.
Those Klich eigenmodes that correspond to small magnitude of $\epsilon$ (and thus contribute the most to the EE) are of particular interest. In our previous paper
\cite{ref:pouranvariyang2} we showed that the Klich eigenmode corresponding to the smallest magnitude of $\epsilon$ (which we call maximally entangled mode (MEM))
contains very useful physical information about the system.

As a simple example, let us consider the case of two spins. We choose one of them to be subsystem $A$ and the other subsystem $B$. In the corresponding fermionic
representation there are two energy levels. Two spins can be in each one of four different states: singlet or one of three triplet states. If the two spins are in
triplet$_{\downarrow\downarrow}$=$\ket{\downarrow\downarrow}$, then by Eq. (\ref{eq:nf}), $N_F=0$ and thus $n^A=n^B=0$, or $\epsilon^A=\epsilon^B=+\infty$. If the two
spins are in triplet$_{\uparrow\uparrow}$=$\ket{\uparrow\uparrow}$, where $N_F=2$ and $n^A=n^B=1$, or $\epsilon^A=\epsilon^B=-\infty$.
Obviously, the two spins are not entangled in these cases. But when two spins are in singlet=$\frac{1}{\sqrt{2}}(\ket{\uparrow\downarrow} -\ket{\downarrow\uparrow})$
(where
there is one fermion in the lower energy level) or triplet$_{\uparrow\downarrow }$=$\frac{1}{\sqrt{2}}(\ket{\uparrow\downarrow} +\ket{\downarrow\uparrow})$ (where
there is
one fermion in the higher energy level) then $N_F=1$ and $n^A=n^B=1/2$, and two spins are entangled.

\section{RSRG and RSRG-x Applied to Random XX model}
\label{RSRG-X}
\begin{figure}
\includegraphics[scale=.55]{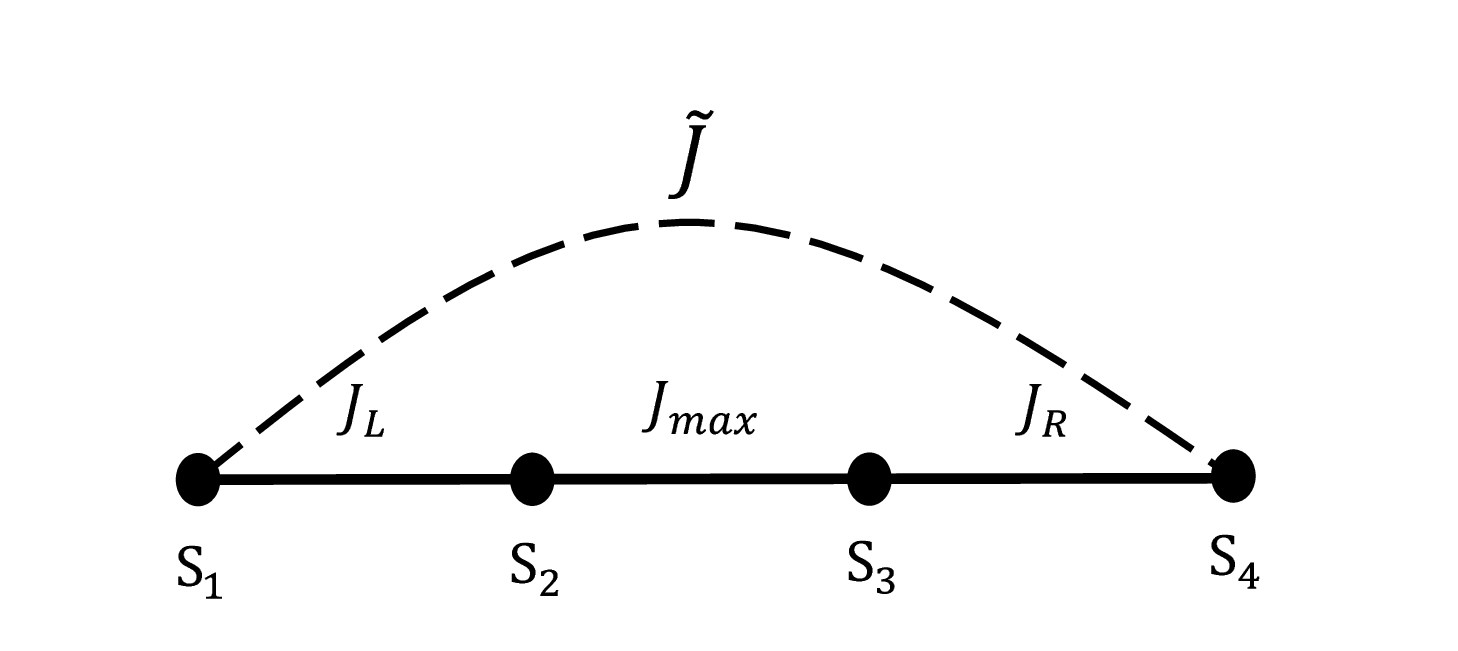}
\caption{\label{fig:rsrg} A schematic plot of one step of real space renormalization group (RSRG/RSRG-X) procedure. Each solid line with coupling $J$ connects two
spins which are depicted by filled circles and the dashed line is an effective coupling $\tilde{J}$ generated by RSRG/RSRG-X.}
\end{figure}

Consider anti-ferromagnetic XX spin-$1/2$ chain with coupling constants $J$'s that are randomly distributed by a distribution function. The approximate ground state
can be obtained using RSRG method as following: at each step of RSRG we choose the biggest $J$ and consider two spins that are coupled by this $J_{max}$ (spins number
$2$ and $3$ in Fig. \ref{fig:rsrg}). We put these two spins to be in the lowest energy state, the singlet state. Since coupling of these two spins is bigger than
neighboring couplings $J_L$ and $J_R$, we treat them as perturbations. First order perturbation vanishes and by using the second order perturbation we obtain an
effective coupling between spins number $1$ and $4$ to be $\tilde{J} \approx \frac{J_L J_R}{J_{max}}$ and we remove spins number $2$ and $3$. This procedure continues
until we form $N/2$ singlet pairs (assuming even $N$). The ground state of the system is approximated as the direct product of these singlet states within RSRG.

\begin{figure}
\includegraphics[width=.5\textwidth]{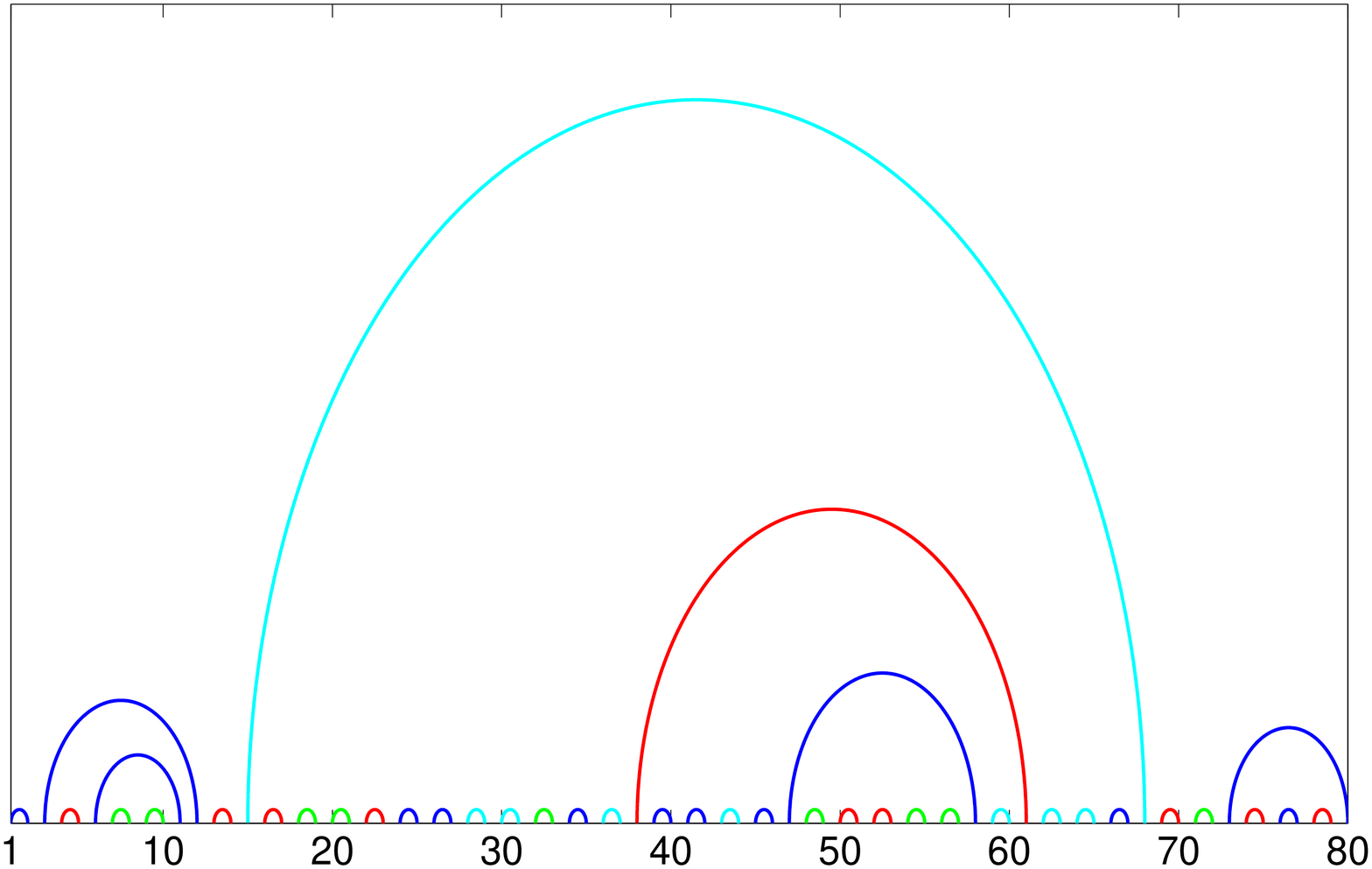}
\caption{\label{fig:rsrg-x}(Color online) An example of spin configuration generated by RSRG-X method to find approximate eigenstate of the XX spin chain corresponding
to a
non-zero temperature. Every two spins are connected by a colored line: singlet (blue), triplet$_{\uparrow\downarrow }$ (green), triplet$_{\uparrow\uparrow }$ (cyan),
triplet$_{ \downarrow\downarrow }$ (red).}
\end{figure}

Excited eigenstates can also be obtained by a modified version of RSRG which is called RSRG-X
\cite{ref:rsrgx}. In this modified method, two spins with the largest coupling constant are chosen to be in one of the four eigenstates with energy $E$ by the
Boltzmann distribution:
\begin{equation}
P=\frac{1}{Z} e^{-E/T}, ~ Z=2+2 \cosh{\frac{J}{2T}},
\end{equation}
namely for each one of the singlet/triplet states, there is a corresponding probability associated with a temperature $T$. The effective coupling $\tilde{J}$ depends
on
which state is chosen for the two spins. Energy, corresponding effective coupling, and the probability to be in the singlet/triplet states are listed in Table
\ref{tb:eigen} \cite{ref:huangmoore}. Note that if two spins are in $\ket{\uparrow\uparrow}$ or $\ket{\downarrow\downarrow}$ state, the effective coupling they mediate
has opposite sign to the other cases. In the end, by using the RSRG-X process, a typical excited eigenstate associated with a temperature is approximated as the
product state of these singlet and triplet pairs. One example of such a state generated by RSRG-X method for $T \neq 0$ is depicted in Fig. \ref{fig:rsrg-x}. In this
figure, singlet and triplet states are plotted in different colors.

\begin{table}
\caption{\label{tb:eigen} Eigenstates and eigenvalues of two spins, the corresponding probability associated with temperature $T$, and effective spins in the RSRG-X
method. $Z=2+2 \cosh{\frac{J}{2T}}$.}
\begin{tabular}{l|c|c|c}
 Eigenstate & Eigenvalue & Probability & \vtop{\hbox{\strut Effective}\hbox{\strut coupling}}   \\
\hline
\hline
singlet= $\frac{1}{\sqrt{2}}[\ket{\uparrow\downarrow} - \ket{\downarrow\uparrow}]$ & $-J/2$ & $\frac{1}{Z} e^{J/2T}$ & $\tilde{J} \approx +\frac{J_L J_R}{J_{max}}$ \\

triplet$_{\uparrow\downarrow }$=$\frac{1}{\sqrt{2}}[\ket{\uparrow\downarrow} + \ket{\downarrow\uparrow}]$ & $+J/2$ & $ \frac{1}{Z} e^{-J/2T}$ &$\tilde{J} \approx
+\frac{J_L J_R}{J_{max}}$ \\

triplet$_{\uparrow\uparrow}$= $\ket{\uparrow\uparrow}$ & $0$ & $\frac{1}{Z}$ & $\tilde{J} \approx -\frac{J_L J_R}{J_{max}}$ \\

triplet$_{\downarrow\downarrow}$= $\ket{\downarrow\downarrow}$ & $0$ & $\frac{1}{Z}$ & $\tilde{J} \approx -\frac{J_L J_R}{J_{max}}$ \\

\end{tabular}
\end{table}

At each step of RSRG-X, effective coupling can be a negative or a positive number ($\tilde{J} \approx \pm \frac{J_L J_R}{J_{max}}$ depending on which of singlet or
triplet states is chosen according to the Boltzmann distribution). After repeating RSRG-X steps many times, probability of having (positive or negative) smaller $J$'s
increases and finally we have the following power law distribution function (we set the cutoff energy to be $1$):

\begin{equation}\label{prob}
P(J) =
\left\{
	\begin{array}{ll}
		\frac{\alpha}{2} |J|^{-1+\alpha}  & \mbox{, if }  |J| \leq 1 \\
		0                                   & \mbox{, otherwise}
	\end{array}
\right.
\end{equation}
with $\alpha$ approaching zero slowly with decreasing energy scale, or disorder strength increases with lowering energy scale. This is the same behavior as in RSRG,
which establishes their asymptotically exact nature. We choose this distribution function in
our exact numerical calculations. In this distribution function small $\alpha$ corresponds to strong disorder regime and large $\alpha$ corresponds to weak disorder
regime.

In the fermionic representation of the Hamiltonian, Eq. (\ref{eq:hamc}), we need to know the distribution of fermions in energy levels corresponding to the eigenstate
obtained in RSRG/RSRG-X method. First we calculate the single particle energy levels by diagonalizing the Hamiltonian in Eq. (\ref{eq:hamc}). Due to the particle-hole
symmetry of the Hamiltonian, energy levels form $\pm E$ pairs. For a system with $N$ sites, there are $N/2$ pairs of energy levels (assuming even $N$).  Each pair of
energy levels corresponds to an effective bond in the RSRG/RSRG-X method. For the ground state of the Hamiltonian, for each step of RSRG, we put the two spins coupled
with the maximum $J$ to be in the ground state and correspondingly we put one fermion in the negative sector of each pair of $\pm E$ and thus all negative energy
levels are filled. The procedure is different in the RSRG-X application for excited eigenstates.
At each step of RSRG-X, number of fermions ($0$, $1$, or $2$) and which energy level(s) we put them in, depend on the state of the two spins (the two spins can be in
singlet or either of triplet).
For $N/2$ pairs
of energy levels, we have $N/2$ RSRG-X steps; for $n$th step, the corresponding pair of energy levels are the $n$th and $(N-n+1)$th energy levels. We put no fermion if
two spins are chosen to be
in $\ket{\downarrow\downarrow}$. If Boltzmann distribution chooses two spins to be in $\ket{\uparrow\uparrow}$, then we put two fermions, one in each of two specified
energy levels. For the case of $\frac{1}{\sqrt{2}}[\ket{\uparrow\downarrow} \pm \ket{\downarrow\uparrow}]$, we put one fermion in $n$th or $(N-n+1)$th level depending
the
randomly chosen state is a ground state or an excited state (by considering the sign of the $J_{max}$).

\section{Results}
\label{results}

With regard to entanglement, predictions of the RSRG-X method are the following. The excited eigenstate is approximated as the product state of the singlet and triplet
states obtained in all
steps. When the system is divided to two subsystems, it may cut singlet and/or triplet states. For each singlet and triplet$_{\uparrow\downarrow}$ crossing the
boundary, there is one entanglement energy $\epsilon$ at exactly $0$ and the rest are $\pm \infty$. In addition, corresponding Klich eigenmode for each $\epsilon=0$
is
highly localized at two spins in singlet or triplet$_{\uparrow\downarrow}$. But since spins in  triplet$_{\uparrow\uparrow}$  and triplet$_{\downarrow\downarrow}$
state are not entangled and do not contribute to the entanglement, Klich eigenmodes do not display the location of spins in these two states.

In this section we analyze the predictions of the RSRG-X method. In first subsection, by looking at the entangled modes, we explain how they represent the entangled
spins. In the subsequent subsection, we examine the predictions of RSRG-X about entanglement energies. For both subsections, we distinguish between strong and weak
disorder limits to see in which limit RSRG-X predictions are accurate.

\subsection{Entangled modes}

In our previous papers \cite{ref:pouranvariyang1, ref:pouranvariyang2} we explained that MEM reveals important physics about the system. For example MEM displays
location of the pair of spins in singlet state that crosses the boundary of subsystem (and thus contribute to the entanglement) in the ground state of the XX chain.
Here, we show that MEM displays the location of spins that are in singlet or triplet$_{\uparrow\downarrow }$ that cross the boundary for the excited states of the
XX chain.

\begin{figure*}
\centering
\subfigure[]{
\includegraphics[width=0.48\textwidth]{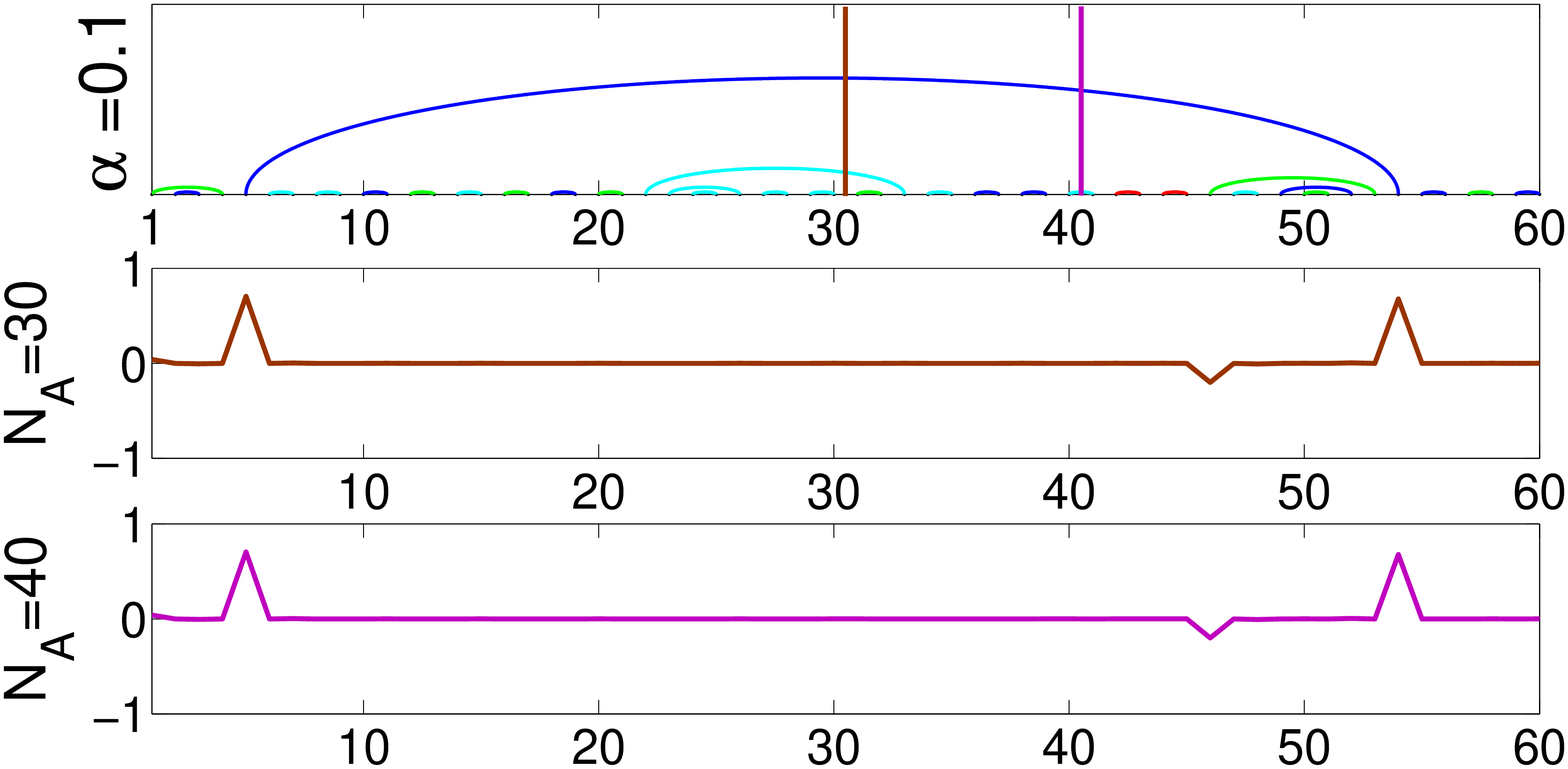}
}
\subfigure[]{
\includegraphics[width=0.48\textwidth]{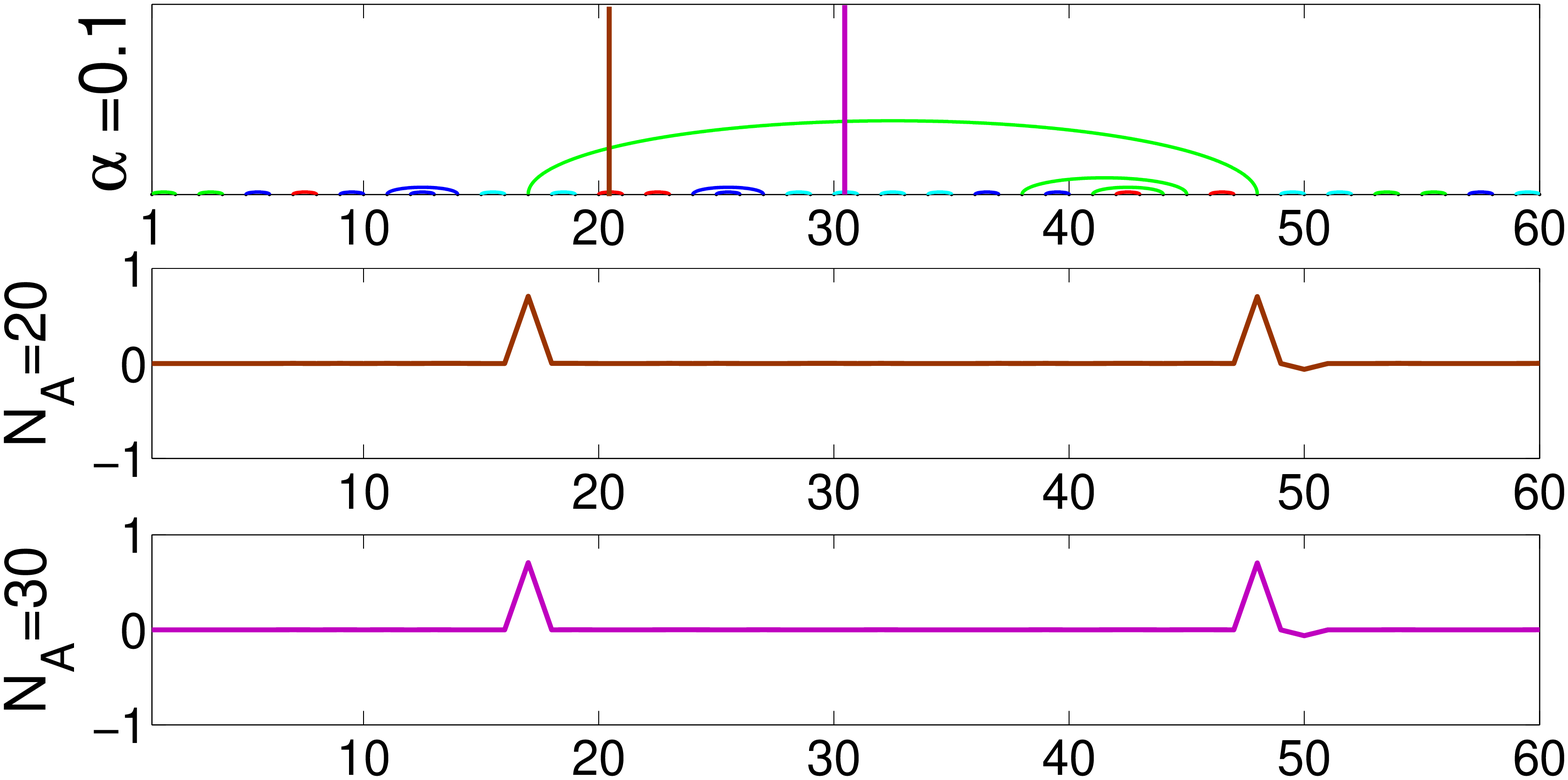}
}
\\
\subfigure[]{
\includegraphics[width=0.48\textwidth]{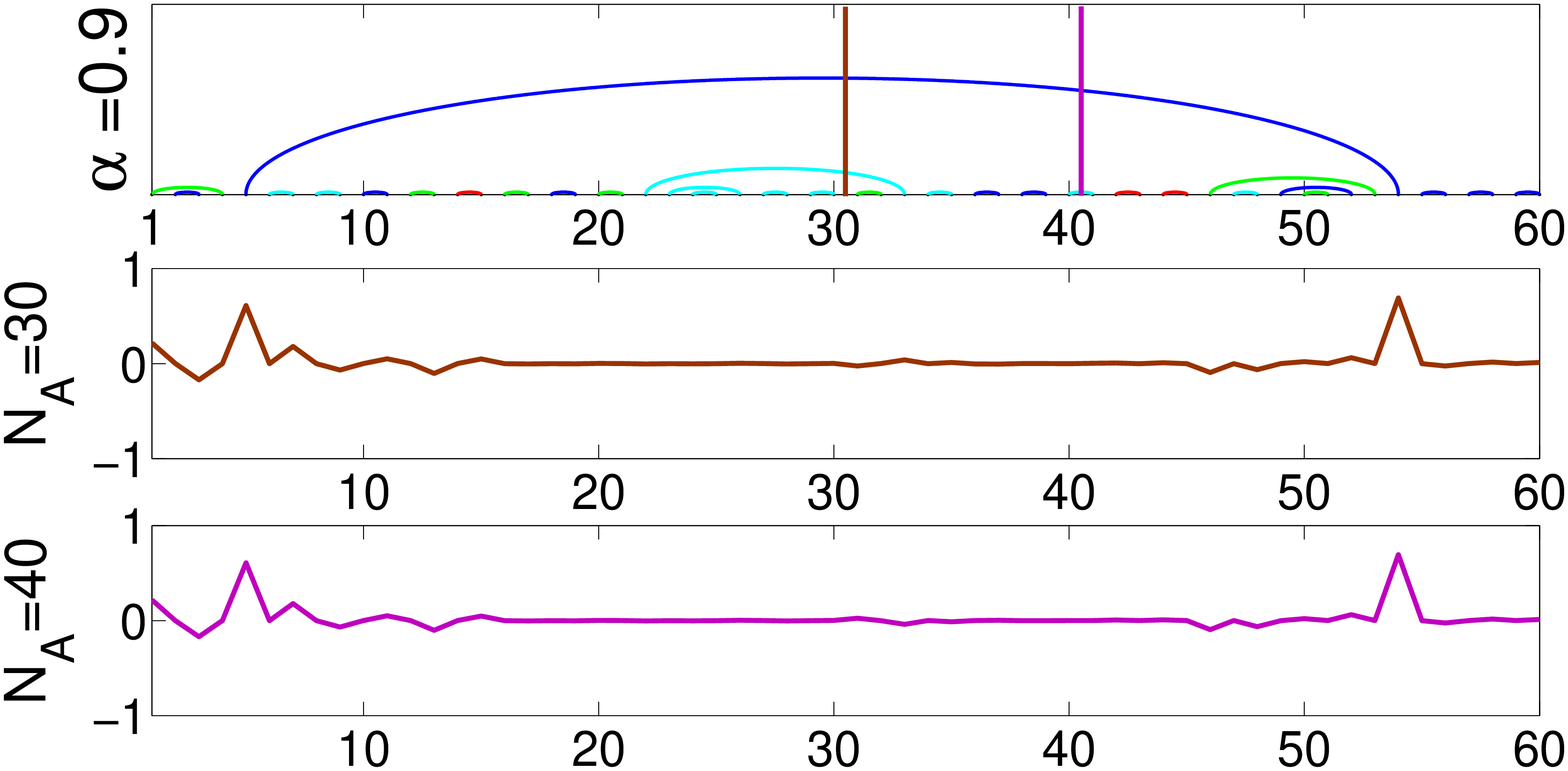}
}
\subfigure[]{
\includegraphics[width=0.48\textwidth]{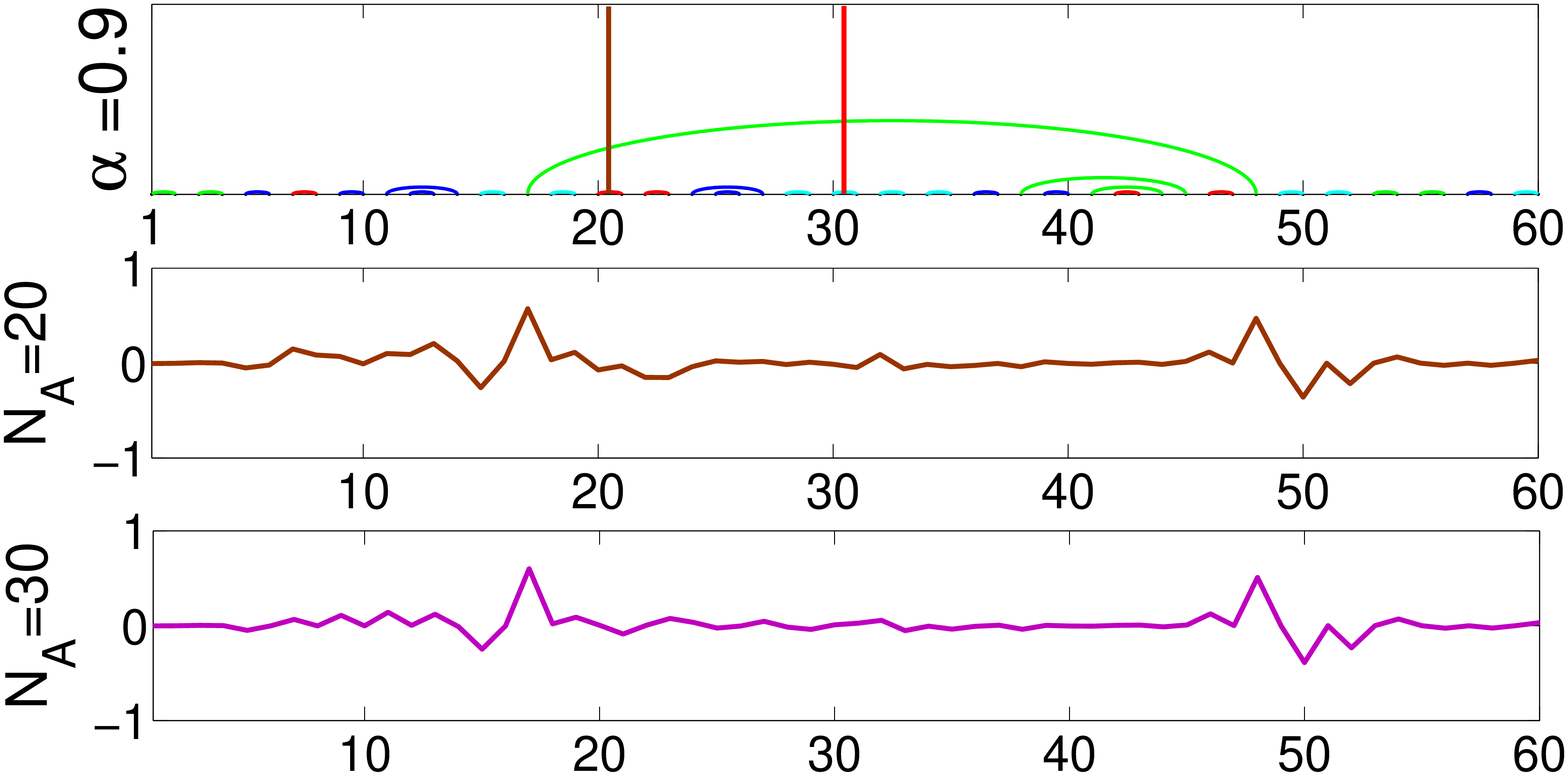}
}
\caption{\label{fig:st19}(color online) Demonstrating how MEM displays the location of entangled spins. In all four plots we set $N=60$ and $T=1$. Subplots (a) and
(b)
correspond to the strong disorder ($\alpha=0.1$); (c) and (d) correspond to the weak disorder ($\alpha=0.9$). In subplots (a) and (c) there is only one entangled
singlet (blue
bond) and in subplots (b) and (d) there is only one entangled triplet$_{\uparrow\downarrow }$ (green bond). First panel of each subplot displays the RSRG-X method.
Each
two spins in singlet/triplet states are connected by a line with a specific color. Singlet (blue), triplet$_{\uparrow\downarrow}$ (green),
triplet$_{\uparrow\uparrow}$ (cyan), and triplet$_{\downarrow\downarrow}$ (red). Second and third panels show the MEM with specified subsystem size $N_A$.}
\end{figure*}

In Fig. \ref{fig:st19}, there are four subplots, by which we want to demonstrate how MEM displays the location of entangled spins. In all of them we set $N=60$ and
$T=1$. There are three panels for each subplot. First panel is the presentation of the RSRG-X where we show each two spins in the singlet/triplet states connected by a
colored line, second and third panels are the MEM associated with the specified partition of the system shown in the plot by a vertical line. In subplot (a), first
panel, we show the RSRG-X of a sample that has only one singlet crossing the boundary and we choose  $\alpha=0.1$ (strong disorder regime). In
subplot (b), we choose a sample that has only one  triplet$_{\uparrow\downarrow }$ crossing the boundary and also with $\alpha=0.1$.  In second panel of both plots we
plot the corresponding MEM. We repeated the same calculations in subplots (c) and (d), but instead we choose $\alpha=0.9$ (weak disorder regime).

Comparing two cases of strong and weak disorder, we observe that: First, for the strong disorder case MEM is highly localized at the location of spins that are
predicted by RSRG-X to form
singlet or triplet$_{\uparrow\downarrow }$, clearly demonstrating its accuracy. For the weak disorder case, while MEM is more extended, we still see that it is
centered around spins that RSRG-X predicts to form singlet or triplet$_{\uparrow\downarrow }$ that cross the boundary. In this case what get entangled are effective
spins renormalized through RSRG-X procedures, and MEM gives their profiles.

Second, to see
that these effective spins are intrinsic of the system and independent of the boundary, we choose a different partition in the third panel of each of the subplots and
we
see that still MEM is localized at the same entangled (effective) spins.
More quantitatively, we calculate overlap between MEM's corresponding to different subsystem sizes in Table \ref{tb:overlap1} and \ref{tb:overlap9}. We see that MEM's
of
case with different subsystem sizes but corresponding to same entanglement configuration are almost identical, although in the week disorder case their overlaps are
slightly
smaller than the strong disorder case. Thus, MEM is a property of the system and it is independent of subsystem boundary.

\begin{table}
\caption{\label{tb:overlap1}Overlap between MEM's corresponding to different subsystem sizes for $\alpha=0.1$ and $\alpha=0.9$, same setting as Fig. \ref{fig:st19},
subplots (a) and (c) respectively. Subsystem size with $N=55$ corresponds to no entanglement. Other subsystem sizes correspond to the same case of one singlet crossing
the boundary.}
\begin{tabular}{cccccc}
\hline
\hline
$\alpha=0.1$ & $N_A=15$ & $N_A=19$ & $N_A=30$ & $N_A=40$ & $N_A=55$ \\
\hline
$N_A=15$ & 1 & 1 & 1 & 1 & $1\times 10^{-16}$ \\

$N_A=19$ & 1 & 1 & 1 & 1 & $1\times 10^{-16}$ \\

$N_A=30$ & 1 & 1 & 1 & 1 & $1\times 10^{-16}$ \\

$N_A=40$ & 1 & 1 & 1 & 1 & $1\times 10^{-16}$ \\

$N_A=55$ & $1\times 10^{-16}$ & $1\times 10^{-16}$ & $1\times 10^{-16}$ & $1\times 10^{-16}$ & 1 \\
\hline
\hline
\end{tabular}
\begin{tabular}{cccccc}
$\alpha=0.9$ & $N_A=15$ & $N_A=19$ & $N_A=30$ & $N_A=40$ & $N_A=55$ \\
\hline
$N_A=15$ & 1 & 0.99 & 0.99 & 0.99 & $0.0004$ \\

$N_A=19$ & 0.99 & 1 & 0.99 & 0.99 & $0.0004$ \\

$N_A=30$ & 0.99 & 0.99 & 1 & 0.99 & $0.0004$ \\

$N_A=40$ & 0.99 & 0.99 & 0.99 & 1 & $0.0004$ \\

$N_A=55$ & $0.0004$ & $0.0004$ & $0.0004$ & $0.0004$ & 1 \\
\hline
\hline
\end{tabular}
\end{table}

\begin{table}
\caption{\label{tb:overlap9}Overlap between MEM's corresponding to different subsystem sizes for $\alpha=0.1$ and $\alpha=0.9$, same setting as Fig. \ref{fig:st19},
subplots (b) and (d) respectively. Subsystem size with $N=5$ corresponds to no entanglement. Other subsystem sizes correspond to the same case of one
triplet$_{\uparrow\downarrow }$ crossing
the boundary.}
\begin{tabular}{cccccc}
\hline
\hline
$\alpha=0.1$ & $N_A=5$ & $N_A=20$ & $N_A=30$ & $N_A=33$ & $N_A=37$ \\
\hline
$N_A=5$ & $1$ & $1\times 10^{-11}$ & $1\times 10^{-11}$ & $1\times 10^{-11}$ & $1\times 10^{-11}$ \\

$N_A=20$ & $1\times 10^{-11}$ & 1 & 0.99 & 0.99 & 0.99   \\

$N_A=30$ & $1\times 10^{-11}$ & 0.99 & 1 & 1 & $0.99$ \\

$N_A=33$  & $1\times 10^{-11}$ & 0.99 & 1 & 1 & $0.99$\\

$N_A=55$ & $1\times 10^{-11}$ & 0.99 & 0.99	 & 0.99 & $1$ \\
\end{tabular}
\begin{tabular}{cccccc}
\hline
\hline
$\alpha=0.9$ & $N_A=5$ & $N_A=20$ & $N_A=30$ & $N_A=33$ & $N_A=37$ \\
\hline
$N_A=5$ & $1$ & $0.004$ & $0.002$ & $0.002$ & $0.002$ \\

$N_A=20$ & $0.004$ &1 & 0.86 & 0.83 & 0.83   \\

$N_A=30$ & $0.002$ & 0.86 & 1 & 0.83 & $0.83$ \\

$N_A=33$  & $0.002$ & 0.83 & 0.83 & 1 & $0.97$\\

$N_A=55$ & $0.002$ & 0.83 & 0.83 & 0.97 & $1$ \\
\hline
\hline
\end{tabular}
\end{table}

\begin{figure}
\includegraphics[width=0.48\textwidth]{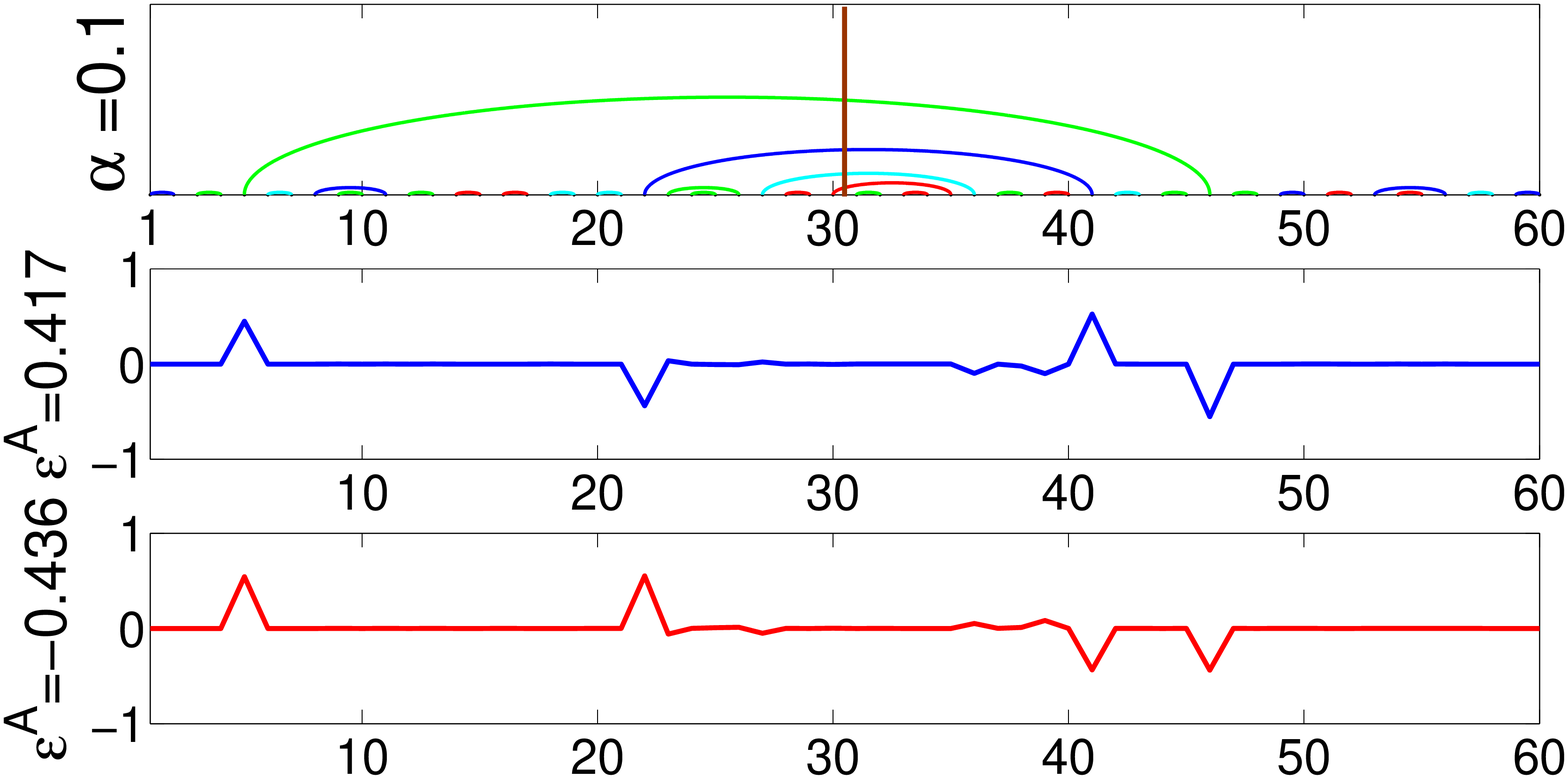}
\caption{\label{fig:s1t1}(color online) An example of having one singlet (blue), one triplet$_{\uparrow\downarrow}$ (green), one triplet$_{\uparrow\uparrow}$
(cyan), and one triplet$_{\downarrow\downarrow}$ (red) cross the boundary. First panel is the configuration of spins making singlet/triplet generated by RSRG-X
method.
Second and third panels are the Klich eigenmode corresponding to the two smallest entanglement energy $\epsilon^A$.}
\end{figure}

It is possible to have cases in which multiple singlet/triplet states cross the boundary of subsystem. We present one example in the following. Fig. \ref{fig:s1t1} is
a case with one singlet, one triplet$_{\uparrow\downarrow}$, one triplet$_{\uparrow\uparrow}$, and one triplet$_{\downarrow\downarrow}$ cross the boundary, and thus
two entangled pairs of spins cross the boundary. In this figure, in which there are two entangled modes, we provide two Klich eigenmodes in second and third panels,
corresponding to the two smallest magnitude of $\epsilon$'s. Both of them are localized at the entangled spins (singlet and triplet$_{\uparrow\downarrow}$) and we also
see that they do
not display the location of the spins in triplet$_{\uparrow\uparrow}$ or triplet$_{\downarrow\downarrow}$ since these two states do not contribute to the
entanglement.

Also, we find that in the strong disorder regime where MEM is localized at two entangled points, sign of the MEM at these two points have a physical information. By
looking at
different examples we find the following rule. Remembering that each singlet/triplet contains specific number of fermions by Eq. (\ref{eq:nf}), we count numbers of
fermions by all singlet and triplet included inside the entangled singlet or triplet$_{\uparrow\downarrow }$ . In the case of entangled singlet if number of fermions
is odd, then both non-zero magnitude of MEM have same sign, otherwise they have opposite signs. The rule is opposite for the case of one entangled
triplet$_{\uparrow\downarrow}$. For example, in Fig. \ref{fig:s1t1}, inside the entangled singlet there are $0$ singlet, $4$ triplet$_{\uparrow\downarrow}$s, $1$
triplet$_{\uparrow\uparrow}$, and $4$ triplet$_{\downarrow\downarrow}$s, so number of fermions included in is $6$ and signs of the Klich eigenmode at the location of
two entangled singlet are opposite. Inside the
entangled triplet$_{\uparrow\downarrow}$ there are $2$ singlets, $7$ triplet$_{\uparrow\downarrow}$s, $5$ triplet$_{\uparrow\uparrow}$s, and $6$
triplet$_{\downarrow\downarrow}$s, so number of fermions included in is $19$ and signs of the Klich eigenmode at the location of two entangled
triplet$_{\uparrow\downarrow}$ are opposite, which are consistent with the proposed rule.

\begin{figure}
\includegraphics[width=.5\textwidth]{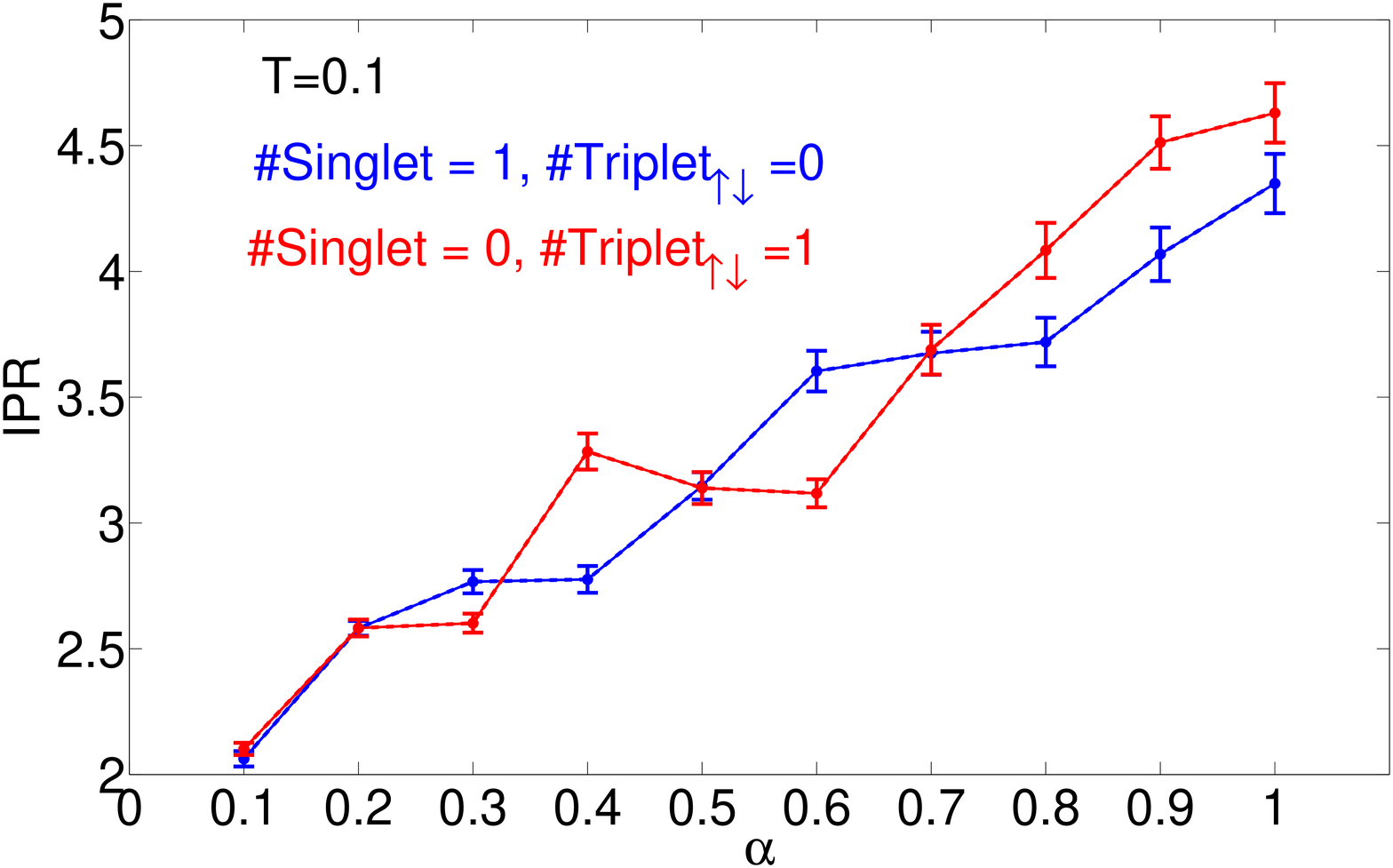}
\caption{\label{fig:IPR_st_alpha} (color online) IPR of the MEM as $\alpha$ changes for a system with $N=100, N_A=50, T=0.1$. Blue line corresponds to the case of
having only one entangled singlet and the red line corresponds to one entangled triplet$_{\uparrow\downarrow }$. IPR becomes bigger as $\alpha$ increases for both
cases. For each $\alpha$, IPR is calculated for $40$ different random $J$'s and for each of $J$'s distributions, thermal averaging is done over $20$ excited states
generated by RSRG-X corresponding to the specific temperature $T=0.1$. Overall, there are $800$ number of data for each $\alpha$.}
\end{figure}

To quantify how much MEM is spread over sites, we use inverse participation ratio (IPR):
\begin{equation}
\text{IPR}=\frac{1}{\sum_i | \psi_i |^4}.
\end{equation}
If IPR is close to $2$, it means that MEM is very localized at the positions of two spins, and when IPR is bigger than $2$, there is an spread of the MEM over a few
sites and IPR$/2$ is the spatial extent of effective spins. In Fig. \ref{fig:IPR_st_alpha}, we plot IPR of the MEM as $\alpha$ changes for two different cases of
having only on singlet or only one triplet$_{\uparrow\downarrow }$ crossing boundary for a non zero temperature $T=0.1$. For both of them we see that in the strong
disorder regime (small $\alpha$) IPR is close to $2$ and this means that MEM is very localized at the position of two entangled spins. This is consistent with RSRG-X
predictions, since we know that in this method (which is asymptotically exact in the strong disorder regime) at each step two spins make a singlet or triplet. As we
approach to the weak disorder regime, IPR increases; meaning that MEM is spread over a few spins and effective spins form entanglement.

\begin{figure}
\includegraphics[width=.5\textwidth]{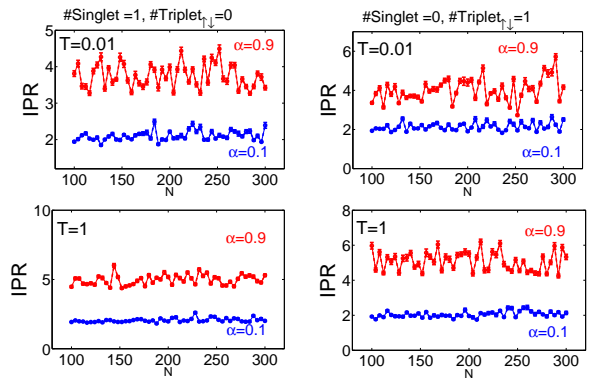}
\caption{\label{fig:IPR_st_N} (color online) IPR of the MEM as $N$ changes between $100$ and $300$ for two cases of a low temperature ($T=0.01$) and a high
temperature
($T=1$). Blue line corresponds to $\alpha=0.1$ and red line corresponds to $\alpha=0.9$. Left (right) panels correspond to one entangled singlet (one
entangled triplet$_{\uparrow\downarrow}$) crossing the boundary. Despite small fluctuations, IPR stays constant as $N$ changes, although for the case of weak disorder,
fluctuations are bigger. For each $N$, IPR is calculated for $40$ different random $J$'s and for each of $J$'s distributions, thermal averaging is done over $20$
excited states generated by RSRG-X corresponding to the specific temperature. Overall, there are $800$ number of data for each $N$.}
\end{figure}

Also, to show that effective spin size is independent from the system size, we provide a plot of IPR of the MEM versus system size $N$ (we set $N_A=N/2$), for two
cases of strong and
weak disorder regime when we have one singlet or one triplet$_{\uparrow\downarrow }$ crossing the boundary in Fig . \ref{fig:IPR_st_N}. Beside small fluctuations, we
see that IPR is constant as system size $N$ changes. This is true for having one singlet or one triplet$_{\uparrow\downarrow }$ crossing the boundary, i.e the
effective spin is a property of the of the MEM.

\begin{figure}
\includegraphics[width=0.5\textwidth]{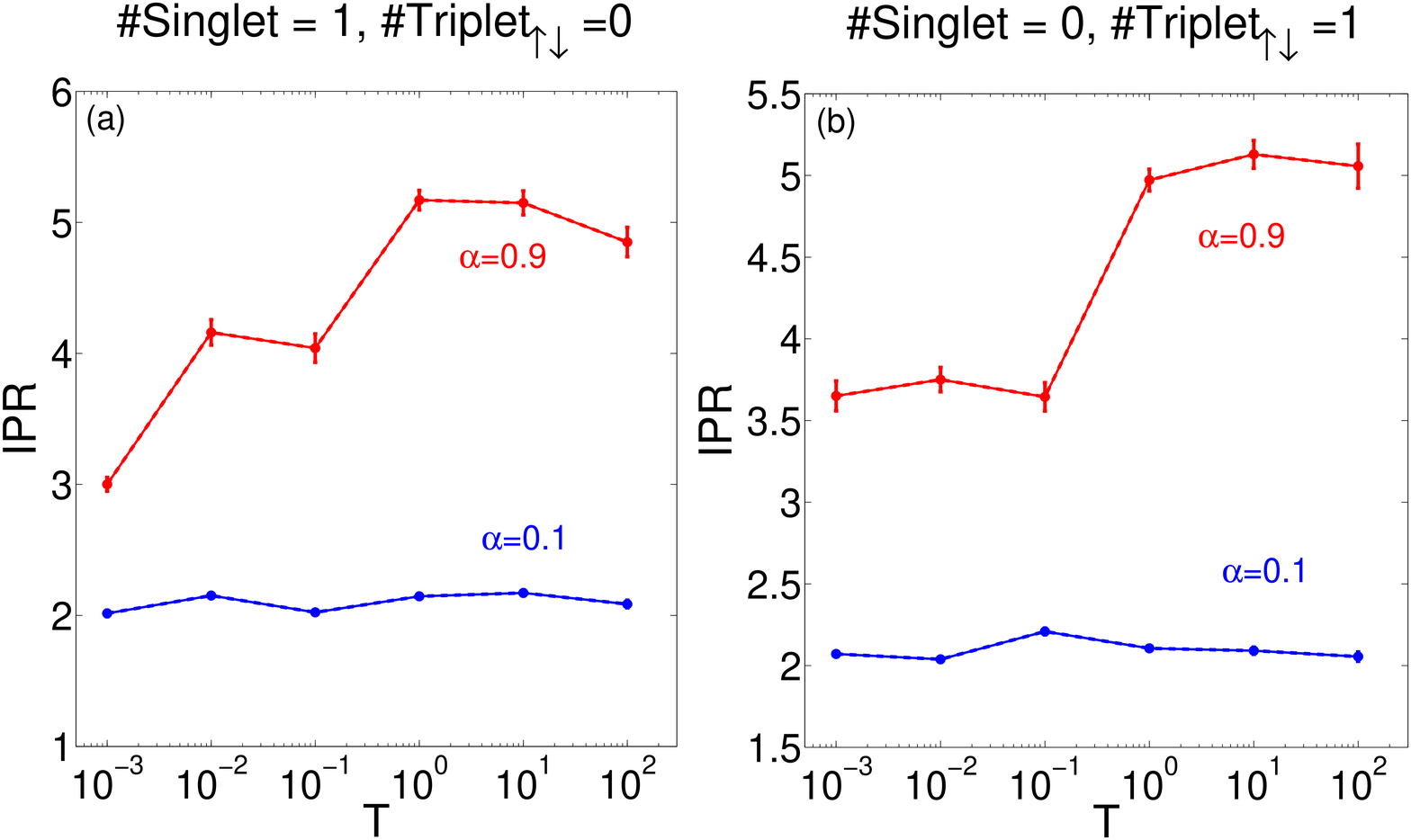}
\caption{\label{fig:IPR_st_T} (color online) IPR of the MEM as temperature changes between $0.001$ and $100$ in log-linear scale for two case of $\alpha=0.1$ (blue
line) and $\alpha=0.9$ (red line). We set $N=100, N_A=50$ for all calculations. In left panel, cases with only one entangled singlet and in right panel cases with only
one
triplet$_{\uparrow\downarrow}$ are considered. For each $T$, IPR is calculated for $40$ different random $J$'s and for each of $J$'s distributions, thermal averaging
is done over $20$ excited states generated by RSRG-X corresponding to the specific temperature. Overall, there are $800$ number of data for each $T$.}
\end{figure}

More importantly, we consider the change of IPR of the MEM as temperature $T$ changes to see how approaching excited states affects the validity of the RSRG-X method.
In Fig. \ref{fig:IPR_st_T} we plot IPR of the MEM for a range of temperatures, from very small $T=0.001$ to a large $T=100$, again for two cases
of having only one singlet or one triplet$_{\uparrow\downarrow }$ crossing the boundary. We see that in both cases IPR is close to $2$ for strong disorder case, i.e.
MEM is highly localized at the locations of two entangled spins even for a highly excited state associated with a high $T$; this is consistent with the predictions of
the RSRG-X method for a finite temperature and thus in the strong disorder regime
RSRG-X method is still valid for excited states corresponding to $T\neq 0$ (for the ground state, $T=0$, we concluded to the same conclusion in our previous paper
\cite{ref:pouranvariyang1}). On the other hand, for the weak disorder case, IPR is always bigger than $2$ and it becomes bigger for higher temperatures, i.e. spatial
extent of the effective spins increases as temperature increases.

\begin{figure}
\includegraphics[width=.5\textwidth]{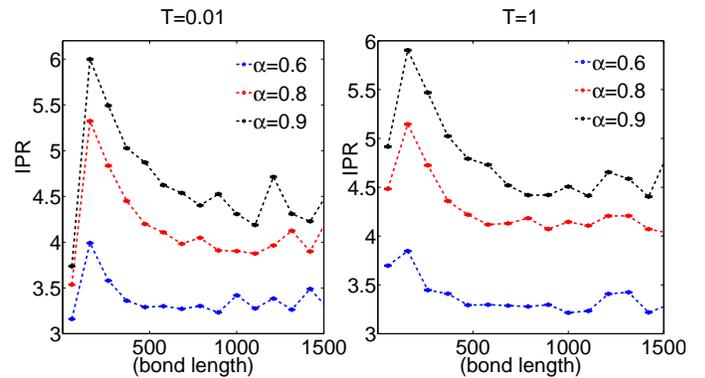}
\caption{\label{fig:IPR_bondlength} (color online) IPR of the MEM versus bond length when system size goes between $100$ and $1500$ with step of $100$. IPR of the MEM
of samples with only one singlet or one triplet$_{\uparrow\downarrow}$ crossing the boundary are considered. IPR is calculated for two temperatures $T=0.01$ (left
panel) and $T=1$ (right panel) and in each panel for three different $\alpha =0.6,0.8,0.9$. For each $N$, IPR is calculated for $40$ different random $J$'s
corresponding to a specific $\alpha$ and for each of $J$'s distributions, thermal averaging is done over $20$ excited states generated by RSRG-X corresponding to the
specific temperature. Range of bond length which is between $1$ and $1500$ is divided to equal segments and for each segment average of the IPR is plotted. For both
small and large temperatures IPR approaches to a constant dependent on the disorder strength $\alpha$. }
\end{figure}

Fig. \ref{fig:IPR_bondlength} illustrates the behavior of the MEM with varying bond length. In this figure, IPR of the MEM of samples with only one singlet or one
triplet$_{\uparrow\downarrow}$ crossing the boundary are considered. IPR of the MEM for two temperatures $T=0.01,1$ and three disorder strengths $\alpha=0.6,0.8,0.9$
is calculated. We see that IPR initially increases, followed by a decrease, and then saturates to a constant for long bond lengths, with the constant depending on
disorder strength. This is consistent with the expectation that the system flows to a fixed point under RSRG-X, as initially it is the bare spins that form singlet or
triplet pairs (hence the small IPR) while renormalization toward effective spins (with growing sizes) occur in the meantime; eventually the renomalization process
saturates, and the size of the effective spins (measured by IPR of MEM) no longer depends on the bond length.

\subsection{Entanglement Energy}
In the RSRG-X treatment, two spins coupled by the strongest bond at a particular step of RG form a singlet or one of the three triplets. For each entangled singlet or
triplet$_{\uparrow\downarrow }$, corresponding entanglement energy is exactly zero. Otherwise entanglement energies are $\pm \infty$ (i.e., no entanglement). In the
following, first we examine this prediction of RSRG-X by looking at the entanglement energies of a specific excited state of one sample (without doing thermal or
sample averaging) with $N=100, N_A=50$ at $T=1$, that has two pairs of entangled spins: on singlet and one triplet$_{\uparrow\downarrow }$. Table \ref{tb:p1_100} and
\ref{tb:p9_100} show the entanglement energies $\epsilon_k^A$'s and the corresponding $n_k^A$ for strong and weak disorder cases respectively. In the strong disorder
case ($\alpha=0.1$), two $\epsilon$'s are very close to zero and beside a few finite $\epsilon$'s, many others are $\pm \infty$. However, for weaker disorder case
($\alpha=0.9$), two smallest $\epsilon$'s are distinguishable from zero and also there are more finite $\epsilon$'s. We see that in the strong disorder case,
entanglement energies are in good agreement with the predictions of RSRG-X.

\begin{figure}
\includegraphics[width=0.5\textwidth]{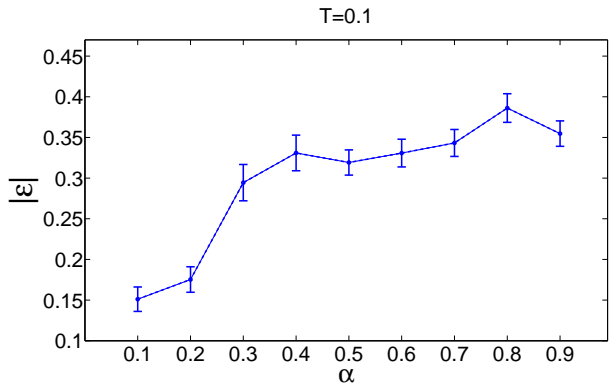}
\caption{\label{fig:epsilon_alpha} (color online) Average of the smallest $\epsilon$ magnitude versus disorder strength. We set $N=100, N_A=50, T=0.1$ and select
samples with one entangled singlet \textit{or} one entangled triplet$_{\uparrow\downarrow }$. For each $\alpha$, $\epsilon$ is calculated for $80$ different random
$J$'s and for each of $J$'s distributions, thermal averaging is done over $40$ excited states generated by RSRG-X corresponding to the specific temperature $T=0.1$.
Overall, there are $3200$ number of data for each $\alpha$.}
\end{figure}

To see the overall behaviour of smallest entanglement energy as disorder strength changes, we plot average of the smallest $\epsilon$ magnitude versus $\alpha$ in
Fig.
\ref{fig:epsilon_alpha}. We set $N=100, N_A=50, T=0.1$ and select samples with one entangled singlet \textit{or} one entangled triplet$_{\uparrow\downarrow }$. We see
that in the strong disorder regime, smallest $\epsilon$ is close to zero and increases as $\alpha$ increases. This is consistent with the expectation that accuracy of
RSRG-X decreases with decreasing randomness strength, just like RSRG.

\begin{figure}
\includegraphics[width=0.5\textwidth]{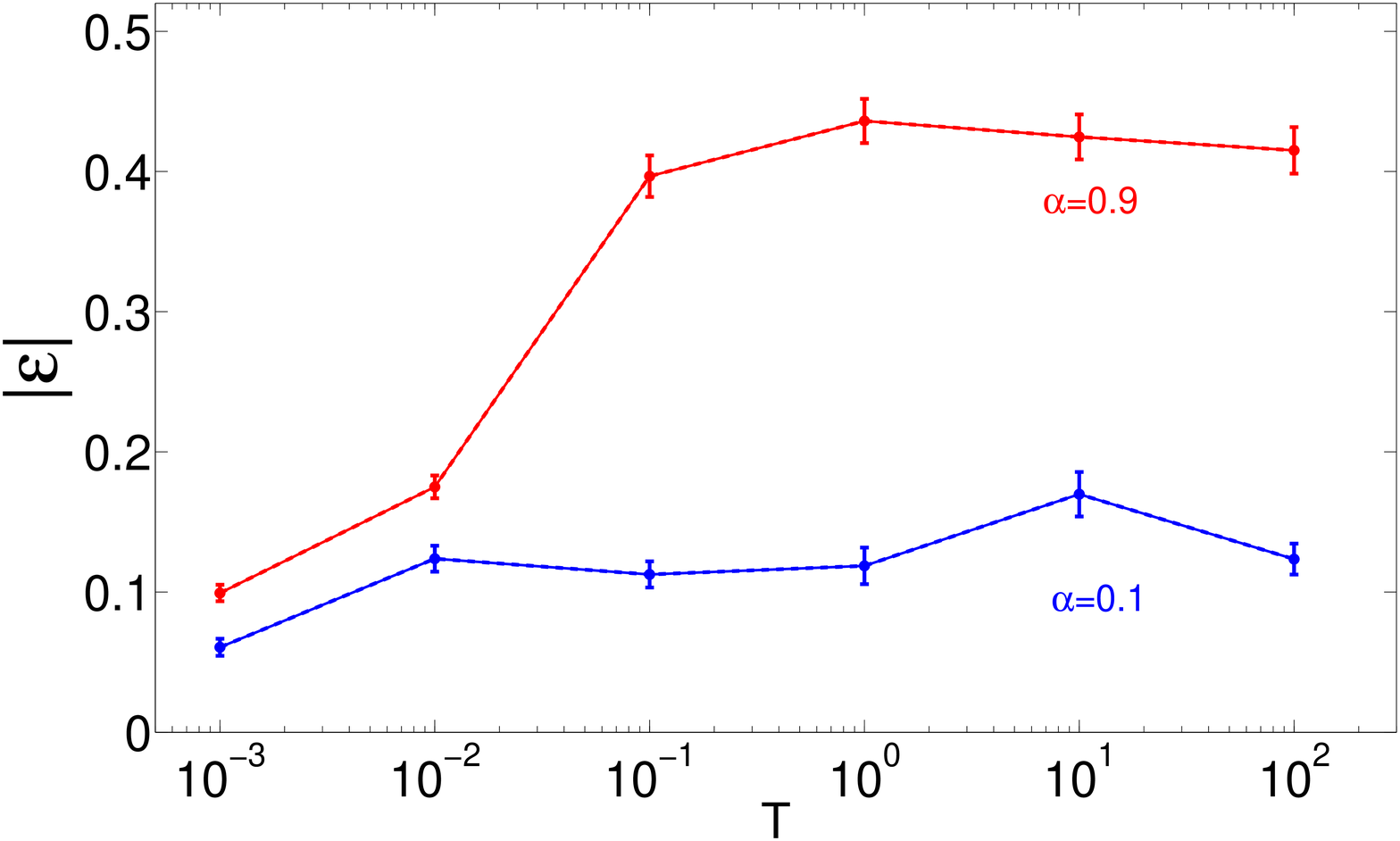}
\caption{\label{fig:epsilon_T} (color online)  Average of the smallest $\epsilon$ magnitude versus temperature $T$. We set $N=100, N_A=50$ and consider samples with
one entangled singlet \textit{or} one entangled triplet$_{\uparrow\downarrow }$. For each
$T$, $\epsilon$ is calculated for $80$ different random $J$'s and for each of $J$'s distributions, thermal averaging is done over $40$ excited states generated by
RSRG-X corresponding to the specific temperature. Overall, there are $3200$ number of data for each $T$.}
\end{figure}

To see how the accuracy depends on temperature/energy density, we plot the average of the smallest $\epsilon$ magnitude versus temperature $T$ for two cases of strong
and weak disorder in Fig. \ref{fig:epsilon_T}. We set $N=100, N_A=50$ and consider samples with one entangled singlet \textit{or} one entangled
triplet$_{\uparrow\downarrow}$. We see that in the strong disorder case ($\alpha=0.1$), $\epsilon$ increases very slowly with temperature, and remains very small for
the entire temperature, attesting to the accuracy of RSRG-X. For the weaker disorder case ($\alpha=0.9$), $\epsilon$ is close to zero for low temperatures and
increases considerably faster with increasing $T$. This indicates the quantitative error of RSRG-X increases with temperature/energy density.

\begin{table*}
\caption{\label{tb:p1_100}Some of $\epsilon_k^A$'s and corresponding $n_k^A$'s for the case of  strong disorder ($\alpha=0.1$). $N=100, N_A=50$, and $T=1$. }
\begin{tabular}{ccccccccccccccc}
\hline
\hline
$\epsilon_k^A$ & $- \infty$ &  $\cdots$  & $-\infty$ & $-8.37$ & $-4\times 10^{-4}$ & $4.4\times 10^{-16}$ & $8.37$ & $23.79$ & $24.69$ & $46.45$ & $66.93$ &$+
\infty$&$\cdots$& $+ \infty$ \\
\hline
$n_k^A $ & $1$ & $\cdots$ & $1$&$0.9997$& $ 0.5001$ & $0.5$& $2\times 10^{-4}$ & $4.6\times 10^{-11}$ & $1.8\times 10^{-11}$ & $\approx 0$ &$\approx 0$ & $ 0$
&$\cdots
$ & $0$ \\
\hline
\hline
\end{tabular}
\end{table*}

\begin{table*}
\caption{\label{tb:p9_100}Some of $\epsilon_k^A$'s and corresponding $n_k^A$'s for the case of  weak disorder ($\alpha=0.9$). $N=100, N_A=50$, and $T=1$. }
\begin{tabular}{ccccccccccccccc}
\hline
\hline
$\epsilon_k^A$ & $- \infty$ &  $\cdots$  & $-36.04$ & $\cdots$  &$-8.46$ &$-7.82$ & $-5.78$ & $-0.270$ & $0.209$ & $5.60$ & $8.08$ & $\cdots$ &$94.9$ & $\infty$\\
\hline
$n_k^A $ & $1$ & $\cdots$ & $\approx 1$&$\cdots$& $ 0.9997$ & $0.9996$& $0.996$ & $0.567$ & $  0.447$ & $0.003$ &$3\times 10^{-4}$ & $\cdots$ &$\approx 0 $ & $0$ \\
\hline
\hline
\end{tabular}
\end{table*}

\section{Concluding Remarks}
\label{conclusion}

We have studied in this paper entanglement energies and entangled modes of highly excited states in a random one-dimensional XX chain, by mapping the system onto a
free fermion model. We found qualitative agreement with predictions based on the newly developed real space renormalization group method for excited states
(RSRG-X)\cite{ref:rsrgx}, lending support to its validity. Furthermore we found RSRG-X is quantitatively accurate in the strong disorder regime. As in the case of RSRG
for ground state there are quantitative errors for weaker randomness, and such error grows with increasing temperature/excitation energy density. This is not
surprising as there are resonances between nearly degenerate excited states not included in the RSRG-X approximation; this source of error does not exist for ground
state.
 Our work is complementary to an earlier study\cite{ref:huangmoore}, which focused on entanglement entropy of the same model. As we demonstrated here and in earlier
 works\cite{ref:pouranvariyang1,ref:pouranvariyang2}, entanglement energies and in particular, entangled modes provide valuable additional information about the
 system.

In a broader context, our work is part of on-going attempt to study entanglement properties of highly excited states. Such properties have been attracting considerable
attention recently, especially in the context of thermalization and many-body localization \cite{mbl}. While most of the existing studies focus on the scaling behavior
entanglement entropy, our work suggests other entanglement measures also deserve investigation.

\acknowledgements

This work was supported by National Science Foundation Grants
DMR-1442366, DMR-1157490 and the State of Florida.

\end{document}